\documentclass[aps,12pt,showpacs]{revtex4}

\usepackage[english]{babel}
\usepackage{amsmath}
\usepackage{amssymb}
\usepackage{amsbsy}
\usepackage{amstext}
\usepackage{graphicx}
\usepackage{subfigure}
\usepackage{axodraw}
\usepackage{pst-node}
\usepackage{verbatim}  

\newcommand{\be}{\begin{eqnarray}}
\newcommand{\ee}{\end{eqnarray}}
\newcommand{\bdm}{\begin{displaymath}}
\newcommand{\edm}{\end{displaymath}}
\newcommand{\ds}{\displaystyle}
\newcommand{\ba}{\begin{array}}
\newcommand{\ea}{\end{array}}
\newcommand{\pa}[1]{\left(#1\right)}
\newcommand{\paq}[1]{\left[#1\right]}
\newcommand{\dpa}{\partial}
\newcommand{\si}{\sigma}
\newcommand{\eps}{\epsilon}
\newcommand{\az}{\mathcal{S}}
\newcommand{\K}{{\bf k}}
\newcommand{\Q}{{\bf q}}

\begin{document}

\title{Classical energy-momentum tensor renormalization via 
effective field theory methods}
\author{Umberto Cannella$^{(1)}$ and Riccardo Sturani$^{(2),(3)}$}
\affiliation{$(1)$ D\'epartement de Physique Th\'eorique, Universit\'e de 
             Gen\`eve, CH-1211 Geneva, Switzerland\\
$(2)$ Istituto di Fisica, Universit\`a di Urbino, I-61029 Urbino, Italy\\
$(3)$ INFN, Sezione di Firenze, I-50019 Sesto Fiorentino, Italy}
\email{e-mail: Umberto.Cannella@unige.ch, Riccardo.Sturani@uniurb.it}

\begin{abstract}

We apply the Effective Field Theory approach to General Relativity, introduced 
by Goldberger and Rothstein, to study point-like and string-like sources in the
context of scalar-tensor theories of gravity.
Within this framework we compute the classical energy-momentum tensor 
renormalization to 
first Post-Newtonian order or, in the case of extra scalar fields, up to the 
(non-derivative) trilinear interaction terms: this allows to write down the 
corrections to the standard (Newtonian) gravitational potential and to the 
extra-scalar potential. 
In the case of one-dimensional extended sources we give an alternative 
derivation of the renormalization of the string tension enabling a re-analysis 
of the discrepancy between the results obtained by Dabholkar and Harvey in one 
paper and by Buonanno and Damour in another, already discussed in the latter.
\end{abstract}


\maketitle

\section{Introduction} 
\label{intro}

We consider in this work the \emph{classical} renormalization of the Energy-Momentum
Tensor (EMT) of fundamental particles and strings due to their interaction with 
long range fundamental fields, including standard gravity.
The gravitational self-energy of a massive body for instance, arises because 
of gravitons' self-interactions, it
can be described as an effective renormalization of the massive body EMT
and it is fully classical having its analog in Newtonian physics.
Such self-interactions, even if they involve point-like particles, are not divergent when 
gravity is present, as on general grounds General Relativity 
imposes a lower limit on the size of massive objects: their Schwartzchild radii.

In the case of one-dimensional extended objects like strings, no horizon analog is present 
and no fundamental lower limit can be imposed on their size: classical contributions to the EMT
due to self-interactions of gravity can (and do indeed) diverge in this case. 
Letting the source size shrink to zero and keeping fixed other physical
parameters like mass and charge (and eventually neglecting gravity), usually 
one encounters infinities, or equivalently,  
physical quantities depending critically on the source size.
Dirac \cite{Dirac} emphasized that the cutoff dependence of the energy of the electromagnetic field 
sourced by an electron can be absorbed by an analog dependence of the bare electron mass, 
to provide a finite, physically observable invariant mass.
However the usual way to consider mass renormalization is by considering
the virtual process of emission and reabsorption of a massless fields, like for
mass renormalization of the electron in standard electrodynamics, rather then
a renormalization of the EMT, i.e. of the particle coupling to gravity, as we are
going to do here.
The above mentioned virtual processes are usually considered in the context of quantum field theory, 
but they show their effects also classically, when heavy, non-dynamical, non-propagating sources
are considered, as we will show.

In order to compute these quantities we make use of the 
the formalism introduced in \cite{Goldberger:2004jt,Goldberger:2007hy}, which is an effective 
field theory (EFT) method borrowed from particle physics, where it originated from 
studying non-relativistic bound state problems in the context of quantum electro- 
and cromo-dynamics \cite{eff-qcd,Luke:1999kz}; for this reason, it has been coined Non 
Relativistic General Relativity (NRGR) (see also \cite{Damour:1995kt} for 
the first appplication of field theory techniques to gravity problems).
Here we apply NRGR in the framework of scalar-tensor theories of gravity 
for computing next-to-leading order corrections to the EMT renormalization,
which in turn define, via the usual Einstein equations, the profile of the graviton
generated by the sources.

An example of such a renormalization has been worked out in 
\cite{BjerrumBohr:2002ks} for point particles
in the GR case and by \cite{Dabholkar:1989jt,Buonanno:1998kx} 
for string-like sources coupled to an extra scalar, the dilaton, and an 
anti-symmetric tensor, the axion. See also \cite{Lund:1976ze,Battye:1994qa,Quashnock:1990wv} for the string sources interacting with axionic and gravitational fields.
We find particularly worth of interest the different analysis performed in 
\cite{Dabholkar:1989jt,Buonanno:1998kx}, leading to apparently 
conflicting result for the string-tension renormalization.
The explanation of the discrepancy is actually given already in \cite{Buonanno:1998kx},
but here we re-analyize such discrepancy with the fresh insight available thanks to NRGR.

The plan of the paper is as follows.
In sec.~\ref{eft} we summarize the basic ingredients of NRGR and set the notation 
for the case at study.
In sec.~\ref{point_part} we apply EFT methods to a model where a scalar
and the standard graviton field mediate long range interactions, to compute the effective EMT of a 
massive body. 
In sec.~\ref{string} we present the analogous computation for a one-dimensional-extended object in four dimensions.
Finally we draw our conclusions in sec.~\ref{conclusion}.

\section{Effective field theory} \label{eft}

We start by describing the basis of NRGR: in doing so we closely follow the thorough  presentation given in \cite{Goldberger:2004jt}, to which we refer for more details, with the exception of the metric signature, as we adopt the ``mostly plus'' convention: 
$\eta_{\mu\nu}\equiv(-,+,+,+)$.

In order to be able to exploit the manifest velocity-power counting, which is 
at the heart of PN expansion, we must first identify the relevant physical 
scales at stake. 
If, for simplicity, we restrict to binary systems of equal mass 
objects it is enough to introduce one mass scale $m$ and two 
parameters of the relative motion, namely the separation $r$ 
and the velocity $v$. It turns out that, up to the very last 
stages of the inspiral, the evolution of the system can be 
modelled to sufficiently high accuracy by non-relativistic dynamics, 
i.e. the leading order potential between the two bodies is 
the Newtonian one. The virial theorem then allows to relate 
the three afore-mentioned quantities according to 
\be
\label{virial}
v^2\sim \frac{G_Nm}{r}
\ee
(where $G_N$ is the ordinary gravitational constant) and tells 
that an expansion in the (square of the) typical three-velocity of the binary 
is at the same time an expansion in the strength of the gravitational 
field.

The compact objects being macroscopic, they can 
be considered fully non-relativistic ($v<<c$) so that from a field 
theoretical point of view, and with scaling arguments in mind,
the binary constituents are non-relativistic particles endowed with 
typical four-momentum of the order $p_\mu\sim (E\sim mv^2, 
{\bf p}\sim m {\bf v})$ (boldface characters are used to denote 
3-vectors). Concerning the motion of the bodies subject to mutual gravitational
potential, it is convenient to consider only the
\emph{potential} gravitons, i.e. those responsible for binding the system as they mediate 
instantaneous interactions: their characteristic four-momentum 
$k_\mu$ will thus be of the order 
\bdm
\label{pot_g}
k_\mu\sim (k^0\sim \frac v{r} , \K\sim\frac 1{r})
\edm
so that these modes are always off-shell ($k_\mu k^\mu\neq 0$).\\
When a compact object emits a single graviton, 
momentum is effectively \emph{not} conserved 
and the non-relativistic particle recoils of a fractional amount roughly 
given by 
$$
\frac{|\delta{\bf p}|}{|{\bf p}|}\simeq \frac{|\K|}{|{\bf p}|}
\simeq \frac\hbar{L}\,,
$$  
where $L\sim mvr$ is the angular momentum of the system:
it is clear that for macroscopic systems such quantity is 
negligibly small.
To summarize, an EFT approach describes massive compact objects 
in binary systems as non-dynamical, background sources 
of point-like type: quantitavely this corresponds to having 
particle world-lines interacting with gravitons. 
The action we consider is then given by 
\be
\az = \az_{EH}+\az_{pp}\,,
\ee
where the first term is the usual Einstein-Hilbert action
\be
\label{az_EH}
\az_{EH}= 2 M^2_{Pl}\int d^4x\sqrt{-g}\ R(g)\,,
\ee
with the Planck mass defined (non canonically) as 
$M^{-2}_{Pl}\equiv 32\pi G_N\simeq 1.2\times 10^{18}$GeV,
and the second term is the point particle action
\be
\az_{pp}=-m\int d\tau = -m\int \sqrt{-g_{\mu\nu} dx^\mu dx^\nu}\,,
\ee
in which $g_{\mu\nu}$ is the metric field that we write as
$g_{\mu\nu}\equiv \eta_{\mu\nu}+ h_{\mu\nu}$\,.
To make the graviton kinetic term invertible, 
one should also include a gauge fixing term like
\be
\az_{gf}=-M^2_{Pl}\int d^4x\ \Gamma_\mu\Gamma^\mu\,,
\ee 
with $\Gamma_\mu\equiv \dpa^\nu h_{\mu\nu}-1/2\,\dpa_\mu h^\nu_\nu$\,.

We now parametrize the metric following \cite{Kol:2007bc}, instead  
of \cite{Goldberger:2004jt}, as
\be
\label{met_nr}
g_{\mu\nu}=\pa{
\ba{cc}
-e^{2\varphi} & -e^{2\varphi}a_j \\
-e^{2\varphi}a_i &\quad e^{-2\varphi}\gamma_{ij}-e^{-2\varphi}a_ia_j\\
\ea
}\,,
\ee
where  $\mu,\nu=0,..,3$ and $i,j=1,2,3$.
We define $\gamma^{ij}$ as the inverse matrix of $\gamma_{ij}$, 
so that $\gamma^{ij}\equiv\pa{\gamma^{-1}}_{ij}$ and 
$a^i\equiv \gamma^{ij}a_j$\,.
It is also useful to introduce 
$\varsigma_{ij}\equiv \gamma_{ij}-\delta_{ij}$ 
(so that $\varsigma^{ij}=\varsigma_{ij}$ to first order) 
and $\varsigma\equiv\varsigma_{ij}\delta^{ij}$.
Then, to quadratic order, the following action for non-canonically 
normalized fields is obtained
\be
\label{az_2}
\left.\az_{EH}\right|_{\rm quadratic}+\az_{gf}=
-\frac{M_{Pl}^2}2\int dt\, d^3{\bf x}
\paq{\dpa_\mu \varsigma_{ij}\dpa^\mu \varsigma_{ij}
- \frac 12\dpa_\mu\varsigma\dpa^\mu\varsigma
+ 8\dpa_\mu\varphi\dpa^\mu\varphi - 2\dpa_\mu a_i\dpa^\mu a_i}\,.
\ee

The non-relativistic parametrization of the metric 
(\ref{met_nr}) allows to write down all the terms 
that do not involve time derivatives in a simple way
\be
\label{az_stat}
\az_{EH}|_{static}=
2M_{Pl}^2\int dt\,d^3{\bf x}\sqrt{-\gamma}\paq{R(\gamma)
-2 \dpa_i\varphi\dpa_j\varphi \, \gamma^{ij} +
\frac 14e^{4\varphi} F_{ij}F_{kl}\gamma^{ik}\gamma^{jl}}\,,
\ee
where $F_{ij}\equiv \partial_ia_j-\partial_ja_i$ is the usual field strength 
tensor.\\
The canonically normalized fields $\sigma_{ij},\phi,A_i$ can be defined as
\be
\ba{ccl}
\sigma_{ij} &\equiv & M_{Pl} \, \varsigma_{ij}\,,\\
\phi &\equiv & 2\sqrt 2M_{Pl} \, \varphi\,,\\
A_i &\equiv & \sqrt 2 M_{Pl} \, a_i\,. 
\ea
\ee
The only interaction term we will need, as it will be explained, is the cubic 
one $\sigma\phi^2$ given by
\be
\left.\az_{EH}\right|_{\si\phi^2}&=&\ds\frac 1{2M_{Pl}}\int dt\,d^3{\bf x}
\paq{\dpa_i\phi\dpa_j\phi\pa{\delta_{ik}\delta_{jl}-
\frac 12\delta_{ij}\delta_{kl}}\si_{kl}}\,.
\ee
The world-line coupling to the graviton thus reads 
\renewcommand{\arraystretch}{1.4}
\be
\label{matter_grav}
\ba{rl}
\az_{pp}=-m\ds \int d\tau &=
\ds -m\int dt\ e^{\phi/(2\sqrt 2 M_{Pl})}
\sqrt{\pa{1-\frac{A_i}{\sqrt 2M_{Pl}}v^i}^2
-e^{-\sqrt 2\phi/M_{Pl}}\gamma_{ij}v^iv^j}  \\
&\simeq \ds -m\int dt\ e^{\phi/(2\sqrt 2 M_{Pl})}\pa{1-\frac 12 v^2 
+ \frac{\phi}{2\sqrt 2 M_{Pl}} -\frac{A_i}{\sqrt 2M_{Pl}}v^i+\ldots}\,.
\ea
\ee
\renewcommand{\arraystretch}{1}

\noindent The propagators we use are given by the following non-relativistic 
expressions, as we are treating the time derivatives in the kinetic terms as 
perturbative contributions,
\renewcommand{\arraystretch}{1.4}
\be
\ba{ccl}
\ds\rnode{s1}\si_{ij}(t,\K)\rnode{s2}\si_{kl}(t',\K')&=&\ds
(2\pi)^3\delta(t-t')\delta^{(3)}(\K-\K')\frac i{\K^2}P_{ij,kl}
\ncbar[nodesep=2pt,angle=-90,armA=4pt,armB=4pt]{s1}{s2}\\
\ds\rnode{a1}A_i(t,\K)\rnode{a2}A_j(t',\K')&=&
\ds(2\pi)^3\delta(t-t')\delta^{(3)}(\K-\K')\frac i{\K^2}\delta_{ij}
\ncbar[nodesep=2pt,angle=-90,armA=4pt,armB=4pt]{a1}{a2}\\
\ds\rnode{f1}\phi (t,\K)\rnode{f2}\phi (t',\K')&=&
\ds (2\pi)^3\delta(t-t')\delta^{(3)}(\K-\K')\frac i{\K^2}
\ncbar[nodesep=2pt,angle=-90,armA=4pt,armB=4pt]{f1}{f2}\\\\
\ea
\ee
\renewcommand{\arraystretch}{1}

\noindent where 
\be
\label{prop_ten}
P_{ij,kl}\equiv\frac 12\pa{\delta_{ik}\delta_{jl}+\delta_{il}\delta_{jk}-
2 \delta_{ij}\delta_{kl}}\,.
\ee

\begin{figure}[htbp]
  \begin{center} 
    \includegraphics[width=.7\linewidth]{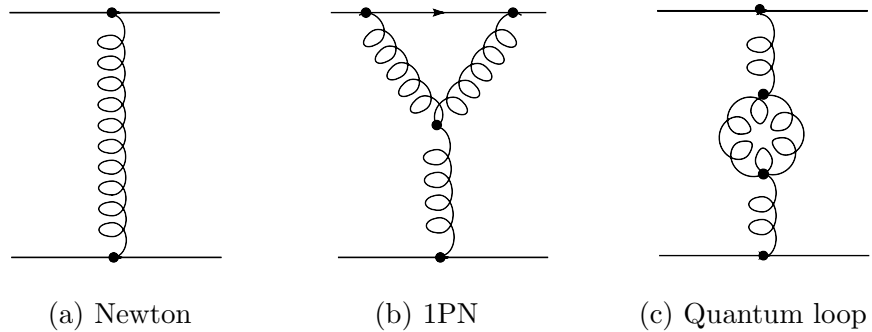}\\
    \hspace{0.5cm}{(a) Newton\hspace{2.3cm} (b) 1PN\hspace{2.0cm} (c) Quantum loop}
    \caption{Contributions to the scattering amplitude of two massive objects. 
      From left to right the diagrams represent respectively the leading Newtonian 
      approximation, a \emph{classical} contribution to the 1PN order and a 
      negligible quantum 1-loop diagram.}
    \label{class_quant}
  \end{center}
\end{figure}

As far as we are only concerned in scaling we can set $k\sim 1/r$, $t\sim r/v$ 
and, by virtue of the virial theorem (\ref{virial}), $m/M_{Pl}\sim \sqrt{Lv}$. 
We can then immediately estimate what are the scalings of the contributions to 
the scattering amplitude of two massive objects: each of the three diagrams 
reported in fig.~\ref{class_quant}, for instance, contributes to such process.
By assigning a factor $[\frac m{M_{Pl}}dt \, d^3\K]$ to a graviton-worldline
coupling not involving velocity, a factor $[\delta(t)\delta^{(3)}(\K)\K^{-2}]$ 
for each propagator, and a factor 
$[\frac{\K^2}{M_{Pl}} dt\,\delta^{(3)}(\K)\pa{d^3\K}^3]$ for a 
three-graviton vertex, the following scaling laws can be associated to the 
different contributions of fig.~\ref{class_quant}:
\renewcommand{\arraystretch}{1.4}
\bdm
\ba{rcl}
(a)&\sim& \ds\pa{\frac m{M_{Pl}}}^2\paq{dt\,d^3\K}^2
\paq{\delta(t)\delta^{(3)}(\K)\K^{-2}}\sim L\,,\\
(b)&\sim& \ds\pa{\frac m{M_{Pl}}}^3\paq{dt \, d^3\K}^3
\paq{\delta(t)\delta^{(3)}(\K)\K^{-2}}^3 
\paq{\frac{\K^2}{M_{Pl}}dt\,\delta^{(3)}(\K)\pa{d^3\K}^3} \sim Lv^2\,,\\ 
(c)&\sim& \ds\pa{\frac m{M_{Pl}}}^2\paq{dt\,d^3\K}^2
\paq{\delta(t)\delta^{(3)}(\K)\K^{-2}}^4
\paq{\frac{\K^2}{M_{Pl}}dt\, \delta^{(3)}(\K)\pa{d^3\K}^3}^2\sim v^4\,.
\ea
\edm
\renewcommand{\arraystretch}{1}

\noindent Even if we are actually dealing with a classical field theory, it is interesting
to give a look at the scalings in powers of $\hbar$.
To restore $\hbar$'s one can apply the usual rule that relates the 
number $\mathcal I$ of internal graviton lines (graviton propagators) to the 
number $\mathcal V$ of vertices and the number $\mathcal L$ of graviton loops
\be
\mathcal L =\mathcal I- \mathcal V+1 \,;
\ee
then, taking into account that each internal line brings a power of $\hbar$ and each 
interaction vertex a $\hbar^{-1}$ from the interaction Lagrangian, the total scaling 
for diagrams where the only external lines are massive particles is $\hbar^{{\cal L}-1}$. 
According to this rule the third diagram of fig.~\ref{class_quant} involves one more 
power of $\hbar$ than the first two.
The diagram with a graviton loop is then suppressed with 
respect to the Newtonian contribution, apart from some powers of $v$, by a 
factor $\hbar/L\ll 1$, whereas the second diagram in fig.~\ref{class_quant} is a 
1PN contribution which does not involve any power of $\hbar$. 
Equivalently one can notice that 
since the massive object is not propagating (there is no kinetic 
term in the Lagrangian for such a source), the 1PN diagram is not a 
loop one.

These scaling arguments remain unchanged when other particles are added, 
like a scalar field, and/or another mass scale is introduced 
\cite{Porto:2007pw}, as we will discuss in sec.~\ref{point_part}, provided
that the virial relation (\ref{virial}) correctly accounts for the leading 
interaction.

\section{Effective energy-momentum tensor in scalar-gravity theory: 
the point particle case}
\label{point_part}

The usual way to obtain an effective action $\Gamma$ out of a fundamental 
action $\az_{fund}$ is by integrating out the degrees of freedom we do not 
want to propagate to infinity 
according to the formal rule 
\be
e^{i\Gamma}\equiv \int{\mathcal D}\Phi\,e^{i\az_{fund}}\,,
\ee
where $\Phi$ denotes the generic field to integrate out.

\begin{figure}[t]
    \includegraphics[width=.7\linewidth]{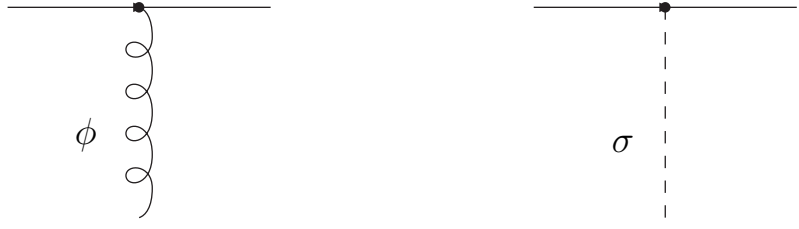}
      \caption{Feynman diagrams describing the gravitational contributions to 
	the effective energy-momentum tensor of a particle at Newtonian level 
	according to the parametrization (\ref{met_nr}) used for the metric.}
  \label{emt_newt}
  \end{figure}
  
In practice this non-perturbative integration is replaced by a perturbative
computation, performed with the aid of Feynman diagrams like those of 
fig.~\ref{class_quant} which shows some contributions to the effective action of two 
particles interacting gravitationally. At lowest order (Newtonian interaction) the diagram in
fig.~\ref{class_quant}(a) represents the term responsible for the Newtonian $1/r$ 
potential between two massive objects. 
Stripping away one of the two external lines in this diagram an amplitude for 
the coupling of a single particle to a graviton is obtained: this amplitude is
linear in the external graviton wave-function and defines the \emph{effective} EMT 
of the particle.
Thus at Newtonian level the two diagrams in fig.~\ref{emt_newt} give the 
following contributions to the effective action
\be \label{g0p}
\ba{rl}
\ds \Gamma^{(0)}=\Gamma_\phi^{(0)}+\Gamma_\si^{(0)}&=\ds
\frac 1{2\sqrt 2 M_{Pl}}\int\phi(x)\paq{T_{00}(x)+T_{ij}(x)\delta_{ij}}\,
d^4x\\
&\ds =\frac{m}{2\sqrt 2M_{Pl}}\int\phi(t,{\bf x_p}(t))\,dt\,,
\ea
\ee
where ${\bf x_p}$ is the three-vector of the position of the source particle 
and use has been made of the Newtonian value of the EMT defined as usual as 
\be
\label{def_emt}
T_{\mu\nu}(x)\equiv \frac{-2}{\sqrt{-g}} \left.
\frac{\delta \az}{\delta g^{\mu\nu}(x)}\right|_{g_{\mu\nu}=\eta_{\mu\nu}}\,.
\ee
Note that the contribution from $\Gamma^{(0)}_\si$ is vanishing as the $\si_{ij}$ 
part of the metric field does not couple directly to a static massive source for 
which $T_{ij}(x)=0$, $T_{00}(x)=m\delta^{(3)}({\bf x}-{\bf x_p})$.

\begin{figure}[t]
  \begin{center}
    \includegraphics[width=.25\linewidth]{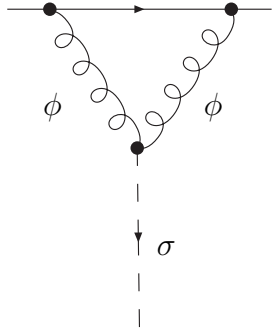}\\
    \caption{Feynman diagram describing the gravitational contribution to the 
      effective energy-momentum tensor of a particle at first post-Newtonian order 
      according to the parametrization used for the metric (\ref{met_nr}).} 
    \label{pp_1PN}
  \end{center}
\end{figure} 

The second diagram in fig.~\ref{class_quant} is a representative  
contribution of the 1PN corrections to the Newtonian potential between two 
particles. Stripping away again one of the two external particle lines the diagram 
showed in fig.~\ref{pp_1PN} is obtained, whose contribution to the 
effective action at next-to-leading order is
\renewcommand{\arraystretch}{1.8}
\be
\label{si_1PN}
\ba{rl}
\ds \Gamma^{(I)}_\si=&\ds \frac 1{M_{Pl}}
\int d^4x\,\sigma_{ij}(x){T^{ij}}^{(I)}(x)=\frac 1{M_{Pl}}
\int dt\,\frac{d^3\Q}{\pa{2\pi}^3} \si_{ij}(t,-\Q){T^{ij}}^{(I)}(t,\Q)
e^{i\Q\cdot {\bf x}_p}\\
=&\ds\frac{m^2}{8M^3_{Pl}}\int dt \frac{d^3\Q}{\pa{2\pi}^3}
\frac{d^3\K}{\pa{2\pi}^3}\frac{k^ik^j-k^iq^j}{\K^2\pa{\K-\Q}^2}
\pa{\delta^l_i\delta^m_j-\frac{\delta_{ij}\delta^{lm}}2}\si_{lm}(t,-\Q)
e^{i\Q\cdot {\bf x}_p}\\
=&\ds \frac{m^2}{2^{10} M^3_{Pl}}\int dt\,\frac{d^3\Q}{(2\pi)^3}
\sigma_{ij}(t,-\Q)\pa{-\delta^{ij}q + \frac{q^iq^j}{q}} 
e^{i\Q\cdot {\bf x}_p}\,,
\ea
\ee
\renewcommand{\arraystretch}{1}

\noindent  where $q\equiv \sqrt{{\bf q}\cdot{\bf q}}$ and we have used eqs.(\ref{kikj}).
The analogous quantity for $\phi$ vanishes as there is no $\phi^3$ vertex, see eq.(\ref{az_stat}).
Incidentally, we note that the EMT obtained from eq.~(\ref{si_1PN}) is transverse, 
consistently with the request that the effective EMT has to be conserved order by order 
(see \cite{Sundrum:2003yt} for an interesting discussion of scalar gravity at interacting level).

Another check of the correctness of our result can be obtained by reconstructing 
the metric out of this effective EMT. The linearized equations of motion for gravity give
\be
\phi(t,\K)=-\frac{1}{\K^2}
\left.\frac{\delta \Gamma_\phi}{\delta\phi(t,\K)}\right|_{\phi=0=\si_{ij}}
\ee
which, using the first of eqs.(\ref{int_k}), allows to compute the metric 
component $\varphi$ according to
\be
\varphi(x) \equiv \frac{\phi(x)}{2\sqrt{2}M_{Pl}}=-\frac m{8 M_{Pl}^2}
\int\frac{d^3\K}{\pa{2\pi}^3} 
\frac {e^{i\K\cdot\pa{{\bf x}-{\bf x_p}}}}{\K^2}=-\frac{G_Nm}r\,,
\ee
where $G_N$ has been reinstated in the final result and 
$r\equiv |{\bf x}-{\bf x_p}|$.
Analogously, for $\varsigma_{ij}$ one has
\be
\varsigma_{ij}(t,\K)=-\frac 1{\K^2}\frac 1{M_{Pl}}P_{ij;kl}
\left.\frac{\delta S}{\delta \si^{kl}(t,\K)}\right|_{\phi=0=\varsigma_{ij}}
\ee
which, again using eqs.(\ref{int_k}), leads to
\be
\ba{rl}
\ds \varsigma_{ij}(t,x)=&\ds P_{ij;kl}\int\frac{d^3\K}{\pa{2\pi}^3} 
\frac{m^2}{2^{10}M_{Pl}^4}\pa{\delta^{kl}k-\frac{k^kk^l}{k}}
\frac 1{\K^2} e^{-i\K\cdot\pa{{\bf x}-{\bf x_p}}}=\\
=&\ds -\frac{\pa{G_Nm}^2}{r^2}\pa{\delta_{ij}-\frac{x_ix_j}{r^2}}\,.
\ea
\ee
Given the metric parametrization (\ref{met_nr}) we obtain 
\renewcommand{\arraystretch}{1.4}
\be
\ba{rcl}
\ds g_{00}&=&\ds -1+\frac{2G_Nm}{r}-2\frac{\pa{G_Nm}^2}{r^2} \\
\ds g_{0i}&=&0 \\
\ds g_{ij}&=&\ds \pa{1+\frac{2G_Nm}r+\frac{\pa{G_Nm}^2}{r^2}}\delta_{ij}+
\frac{\pa{G_Nm}^2}{r^2}\frac{x^ix^j}{r^2}
\ea
\ee
\renewcommand{\arraystretch}{1}

\noindent which is the Schwarzschild metric to 1PN order in the harmonic gauge,
see \cite{BjerrumBohr:2002ks}.

Let us now consider an extra degree of freedom with respect to ordinary 
gravity, that is a massive scalar field $\psi$ whose action is given by
\be
\az_{\psi}= -\frac 12\int d^4x\sqrt{-g}  \paq{g^{\mu\nu}
\dpa_\mu\psi\,\dpa_\nu\psi +m_\psi^2\psi^2+ \lambda \psi^3} \;,
\ee 
where a cubic self-interaction has been allowed. The interaction with the 
gravitational field $\si_{ij}$, embodied by the trilinear term $\psi\psi\si$, 
can be derived from the kinetic term, namely
\be
\label{pps}
\left.\az_\psi\right|_{\psi\psi\si} = \frac 1{2M_{Pl}}\int dt\,d^3{\bf x}\, 
\dpa_i\psi \dpa_j\psi \pa{\sigma^{ij}-\frac 12\delta^{ij}\sigma}\,.
\ee
There are no trilinear terms such as $\phi\psi\psi$ 
or $\phi\phi\psi$ because of the specific metric parametrization we chose
(\ref{met_nr}).
The field $\psi$ is assumed to couple to matter in a metric type
in analogy with (\ref{matter_grav}):
\bdm
\az'_{pp}=-me^{\alpha \psi/\pa{2\sqrt 2M_{Pl}}}\int d\tau \quad ,
\edm
for some dimensionless parameter $\alpha$. Therefore the tree-level coupling 
of $\psi$ to matter at lowest order is very similar to the diagram on the left of 
fig.~\ref{emt_newt}:
\be
\Gamma^{(0)}_\psi=\frac{\alpha \,m}{2\sqrt 2M_{Pl}}\int\psi(t,{\bf x_p}(t))
\, dt\,.
\ee 

\begin{figure}[t]
  \begin{center}
    \includegraphics[width=.25\linewidth]{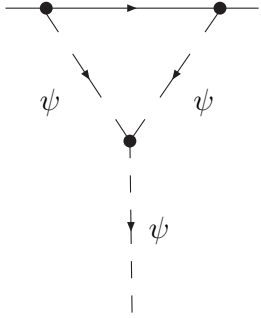}
    \caption{Feynman diagram representing the self-interaction contribution of  
      the massive scalar field $\psi$ to the energy-momentum tensor of a particle 
      at next-to-leading order.}
    \label{3psi}
  \end{center}
\end{figure}

At next-to-leading order we have two possible contributions. The first comes from 
a diagram like that of fig.~\ref{pp_1PN} where the two $\phi$'s  are replaced with 
two $\psi$'s: the amplitude is almost the same as eq.~(\ref{si_1PN}), apart from 
an extra factor $\alpha^2$. 
The second contribution comes from the cubic $\psi$ self-interaction, depicted in the 
diagram of fig.~\ref{3psi}:
\be
\Gamma^{(I)}_\psi=\frac{\lambda m^2\alpha^2}{64\pi M^2_{Pl}}
\int dt\int\frac{d^3\Q}{(2\pi)^3}e^{i{{\bf q}\cdot{\bf x_p}}}
\psi(t,\Q) \frac{1}{q}\arctan{\left(\frac{q}{2m_\psi}\right)}\,.
\ee
Note that at high momentum transfer ($q\gg m_\psi$) the integrand goes as $q^{-1}$, 
whereas in the gravity case (\ref{si_1PN}) we had $T_{ij}^{\si}(q)\propto q$:
this difference leads to an effective potential due to the $\psi$mediation which has a 
logarithmic profile, rather than the $1/r^2$ behavior typical of 1PN terms in 
Einstein gravity derived in \cite{Porto:2007pw}; at low momenta ($q\ll m_\psi$) 
the Yukawa suppression takes place as usual.

\section{Effective energy-momentum tensor: string}
\label{string}

In the case of a one-dimensional extended source we consider the 
Nambu-Goto string with action $\az_s$ given by
\be
\az_s=\mu \int_\Sigma\sqrt{-\gamma}\ e^{\alpha\Phi/\pa{\sqrt 2M_{Pl}}}\,
d\tau d\si-
\frac{\beta\mu}{2\sqrt 2M_{Pl}}\int_\Sigma \dpa_\alpha x^\mu\dpa_\beta x^\nu 
\eps^{\alpha\beta} B_{\mu\nu}\,d\tau d\si\,,
\ee
where $\gamma\equiv {\rm det} \gamma_{\alpha\beta}$, with 
$\gamma_{\alpha\beta}\equiv \dpa_\alpha x^\mu\dpa_\beta x^\nu g_{\mu\nu}$,
 $x^\mu$ are coordinates in the 4-dimensional space, $\si$ and $\tau$ are the 
 coordinates on the world-sheet $\Sigma$ spanned by the string in its temporal 
 evolution.
Such an action describes a fundamental string interacting with gravity via
a string tension $\mu$, with a scalar field $\Phi$ through a coupling 
$\alpha\mu/(\sqrt 2M_{Pl})$ and with the antisymmetric tensor $B_{\mu\nu}$ 
through the coupling $\beta\mu/(2\sqrt 2M_{Pl})$. In this notation a supersymmetric 
string corresponds to $\alpha=\beta=1$.\\ 
The convention for indeces is the following:
$\alpha,\beta$ denote the two directions parallel to the world-sheet while 
$\mu,\nu,\ldots$ are generic 4-dimensional indeces, then Latin 
letters $i,j,\ldots$ denote 3-space indeces and we will use $a,b$ or $c$ to denote 
the (two) spatial dimensions orthogonal to the string.\\
The action $\az_f$ determining the dynamics of the fields is
\be
\label{az_bulk}
\az_f=\int d^4x \sqrt{-g} \paq{2M_{Pl}^2R -\frac 12\pa{\dpa\Phi}^2-
\frac 1{12}e^{-\sqrt 2\alpha\Phi/M_{Pl}}H_{\mu\nu\rho}H^{\mu\nu\rho}}\,,
\ee
where $H_{\mu\nu\rho}\equiv\dpa_\mu B_{\nu\rho}+\dpa_\rho B_{\mu\nu}+
\dpa_\nu B_{\rho\mu}$. The only new propagator we will need with respect to 
the point-particle study is
\be
\label{prop_b}
\rnode{b1}B_{\mu\nu}(t,{\bf k})\rnode{b2}B_{\rho\si}(t',{\bf k'})=\frac 12
\pa{\eta_{\mu\rho}\eta_{\nu\si}-\eta_{\mu\si}\eta_{\nu\rho}}
\pa{2\pi}^3\delta(t-t')\delta^{(3)}(\K-\K')\frac i{\K^2}\,.
\ncbar[nodesep=2pt,angle=-90,armA=4pt,armB=4pt]{b1}{b2}
\ee
Analogously to diagrams in fig.~\ref{emt_newt}, the effective action 
for the linear coupling to the string source of the fields 
$\phi$, $\si_{ij}$, $\Phi$ and $B_{\mu\nu}$ is 
$\Gamma^{(0)}=\Gamma^{(0)}_{\phi}+\Gamma^{(0)}_{\si_{ij}}+\Gamma^{(0)}_{\Phi}+
\Gamma^{(0)}_{B_{\mu\nu}}$ with 
\renewcommand{\arraystretch}{1.4}
\be \label{gammazero}
\ba{rcl}
\Gamma^{(0)}_\phi &=&\ds\int\varphi\pa{T_{00}+T_{ij}\delta^{ij}}\,d^4x=0\,,\\
\Gamma^{(0)}_{\si_{ij}} &=&\ds\int\varsigma_{ij}T^{ij}\,d^4x=
-\frac\mu{M_{Pl}} \int_\Sigma \si_{11}(x(\tau,\si))\,d\tau d\si\,,\\
\Gamma^{(0)}_\Phi &=&\ds\frac{\alpha}{2\sqrt 2M_{Pl}}
\int\Phi\pa{T_{00}-T_{ij}\delta^{ij}}\,d^4x=
\frac{\alpha\mu}{\sqrt 2M_{Pl}}\int_\Sigma \Phi(x(\tau,\si))\,d\tau d\si\,,\\
\Gamma^{(0)}_{B_{\mu\nu}} &=&\ds\frac{\beta\mu}{2\sqrt 2M_{Pl}}
\int_\Sigma\partial_\alpha x^\mu\partial_\beta x^\nu\eps^{\alpha\beta}
B_{\mu\nu}\,d\tau d\si = \frac{\beta\mu}{\sqrt 2M_{Pl}} 
\int_\Sigma B_{01}(x(\tau,\sigma))\,d\tau d\si\,,
\ea
\ee
\renewcommand{\arraystretch}{1} 

\noindent where use has been made of the explicit parametrization of a static string: 
$x^0=\tau$, $x^1=\si$, and of the definition (\ref{def_emt}) for the string EMT 
$T^s_{\mu\nu}$ giving
\be
\label{emt_s}
T_{\mu\nu}^s={\rm diag}(\mu,-\mu,0,0)\delta^{(2)}(x^a)\,.
\ee

\begin{figure}[t]
  \begin{center} 
    \includegraphics[width=.95\linewidth]{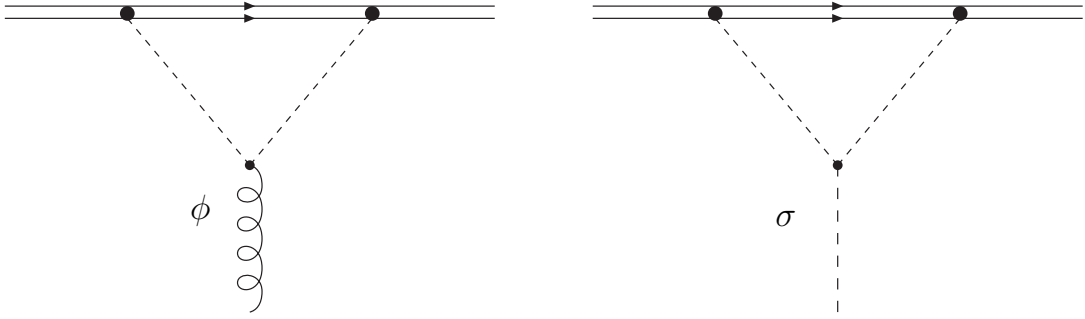}
    \caption{Diagrams reproducing the coupling to $\phi$ (curly line) and to 
     $\sigma_{ij}$ (long-dashed), or the effective energy-momentum tensor, of 
      a string at next to lowest order in interaction. The diagram on the left
      vanishes (see discussion in the text).}
    \label{str_1PN}
  \end{center}    
\end{figure}

Following the same reasoning as in sec.~\ref{point_part}, the contributions to the 
renormalization of the effective EMT due to the 
dilaton and the antisymmetric tensor interaction can be computed, see fig.~\ref{str_1PN}. 
We thus restrict to those trilinear interaction terms involving a graviton 
field, either a $\phi$ or a $\si$, as an external line 
(in a completely analogous way the renormalization of the $\Phi$ and 
$B_{\mu\nu}$ coupling could be computed).
We then have:
\be
\label{s_s3}
\ba{rl}
\ds\az_{3}=\frac 1{2M_{Pl}}&\ds\int dt\,d^3{\bf x} \left\{
\frac 12\paq{\partial_i\Phi\partial_j\Phi
\pa{\delta^{il}\delta^{jm}-\frac 12\delta^{ij}\delta^{lm}}
\sigma_{lm}}+\right.\\
&\ds\left.\!\!\!\!\!\frac 12\paq{\partial_iB_{01}\partial_jB_{01}
\pa{\delta^{il}\delta^{jm}+\delta^{ij}\delta^{l1}\delta^{m1}
-\frac 12\delta^{ij}\delta^{lm}
+\delta^{il}\delta^{j1}\delta^{m1}+\delta^{jm}\delta^{i1}\delta^{l1}}
\si_{lm}}
\right\} \,,
\ea
\ee
where we have specified the antisymmetric tensor polarization indices to "$01$" ,
as this is the only polarization involved in this interaction, and omitted rewriting 
the terms coming from the pure gravity sector, i.e. $\si^3$ and $\phi^2\si$, because 
they read the same as in (\ref{az_stat}).

The diagram on the left in fig.~\ref{str_1PN} is actually vanishing because 
no $\phi$ can attach directly to the string and no trilinear 
term with only one $\phi$ is present in the action (\ref{az_bulk}), as it can be 
seen from (\ref{az_stat}) or (\ref{s_s3}): this implies that the relation 
$T_{00}=-T_{ij}\delta^{ij}$ holds also at next-to-leading order.
We are thus left with the diagram on the right in fig.~\ref{str_1PN}, where the 
particles propagating in the internal dashed lines can be either two dilatons or two 
antisimmetric tensors or two gravitons of the type $\si_{ij}$.
The contribution to $\Gamma^{(I)}_{\si_{ij}}$ from the diagram involving two 
dilatons is 
\renewcommand{\arraystretch}{1.6}
\be
\label{str_1pn}
\ba{rcl}
\ds \Gamma_{\si\Phi\Phi}^{(I)}&=&-\ds\frac{\mu^2\alpha^2}{8M_{Pl}^3}
\int d\tau\frac{d^2q}{\pa{2\pi}^2}e^{-iq_ax_s^a}
\pa{\delta^{ai}\delta^{bj}-\frac 12\delta^{ab}\delta^{ij}}\si_{ij}(\tau,q)
\int \frac{d^2k}{\pa{2\pi}^2}\frac{k_ak_b-k_aq_b}{k^2\pa{k-q}^2}=\\
&=&-\ds\frac{4G_N\mu^2\alpha^2}{M_{Pl}}\int d\tau \frac{d^2q}{\pa{2\pi}^2}
e^{-iq_a x^a_s}\pa{C\delta^{ab}-\frac{q_aq_b}{q^2}}
\pa{\delta^{ai}\delta^{bj}-\frac 12\delta^{ab}\delta^{ij}}\si_{ij}(\tau,q)=\\
&=&\ds \frac{4G_N\mu^2\alpha^2}{M_{Pl}}\int d\tau\frac{d^2q}{\pa{2\pi}^2}
e^{-iq_a x^a_s}
\paq{\pa{-\frac 12\delta^{ab}+\frac{q^aq^b}{q^2}}\si_{ab}(\tau,q)+
\pa{C-\frac 12}\si_{11}(\tau,q)}\,,
\ea
\ee
\renewcommand{\arraystretch}{1}

\noindent with $C$ a divergent quantity, coming from the last integration in 
the first line, whose value can be read from eq.(\ref{kikj_2})
\be
\label{div_const}
C=\lim_{\eps\to 0}\ -\frac 1\eps
\paq{1+\frac \eps 2\pa{\gamma-2+\log\paq{q^2/(4\pi)}+o(\eps)}}\,; 
\ee
here dimensional regularization has been used, as this entry of the effective
EMT is expected to be (logarithmically) UV divergent, see e.g. 
\cite{Dabholkar:1989jt,Buonanno:1998kx}. 
Note that the divergent constant only enters the $T_{11}$ component of the 
effective EMT.

For the $B_{\mu\nu}$ interaction a similar result is obtained 
\be
\Gamma_{\si BB}^{(I)}=\frac{4G_N\mu^2\beta^2}{M_{Pl}}
\int d\tau\frac{d^2q}{\pa{2\pi}^2}e^{-iq_ax^a_s}
\paq{\pa{\frac 12\delta^{ab}-\frac{q^aq^b}{q^2}}\si_{ab}(\tau,q)+
\pa{C-\frac 12}\si_{11}(\tau,q)}\,.
\ee

The contribution to the 1PN effective action due to purely gravitational process, 
i.e. by the diagram on the right of fig. \ref{str_1PN} with three $\si$'s, can be 
computed by making use of the three graviton point function:
\be
\langle\si_{11}(k_1)\si_{11}(k_2)\si_{ij}(q)\rangle=
-\frac 12\delta^{(3)}(k_1+k_2+q) q^2\delta_{i1}\delta_{j1}\,,
\ee
which has been obtained thanks to the Feyncalc tools \cite{Mertig:1990an} for 
Mathematica; the result is
\be
\label{gamma3}
\ba{rl}
\Gamma^{(I)}_{\si\si\si}=& \ds -\frac{\mu^2}{4M^3_{Pl}}
\int d\tau\frac{d^2q}{\pa{2\pi}^2}e^{-iq_a x^a_s}
\delta^{a1}\delta^{b1}\sigma_{ab}(\tau,q)
\int \frac{d^2k}{\pa{2\pi}^2}\frac{q^2}{k^2(k-q)^2} \\
&=\ds \frac{8G_N\mu^2}{M_{Pl}} 
\int d\tau\frac{d^2q}{\pa{2\pi}^2}e^{-iq_a x^a_s}D\,\sigma_{11}(\tau,q) \quad,
\ea
\ee
where $D$ is a divergent constant, again entering the $T_{11}$ component only, 
given by 
\be
\label{nuovoD}
D=\lim_{\epsilon\to 0} \frac 1\epsilon\paq{1+\frac \epsilon 2
\pa{\gamma +\ln \paq{q^2/\pa{4\pi}}}}\,.
\ee
 
The conserved effective EMT is thus given by the sum of the three 
contributions just calculated and reads
\be
\label{eemt_s}
{T_{ij}}^{(I)}(q)&=&4 G_N\mu^2\pa{
\ba{cc}
\pa{2-\alpha^2-\beta^2}D+\alpha^2+\beta^2 & 0\\
0 & \ds \pa{\alpha^2-\beta^2}\pa{-\frac{\delta_{ab}}2+\frac{q_aq_b}{q^2}}\\
\ea}
\ee
together with $T_{00}=-T_{11}$ and $T_{0i}=0$\,.
The coordinate space counterpart of (\ref{eemt_s}) is reported in the 
Appendix.

We note that in the directions orthogonal to the string the EMT is still 
vanishing for $\alpha^2=\beta^2$, thus preserving the no-force
condition valid for supersymmetric strings of the same type (charge).
The divergent part of the entry $T_{11}$ is also vanishing in the supersymmetric 
case due to a cancellation among the different terms: therefore, the 
\emph{superstring tension}, given by $T_{11}$, does not receive divergent contribution.
This confirms the result of Dabholkar and Harvey \cite{Dabholkar:1989jt} 
obtained through the analysis of the EMT's
on the (linearized) GR solution around a string.

\begin{figure}[t]
  \begin{center}
    \includegraphics[width=0.3\linewidth]{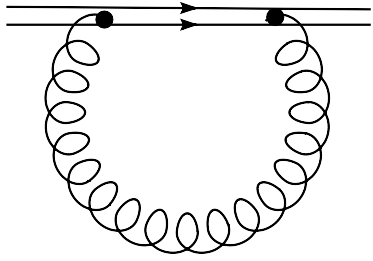}\\
    \caption{Feynman diagram representing the string tension renormalization as computed 
      in \cite{Buonanno:1998kx}. The internal wavy line stands for all possible fields 
      interacting with the string: dilaton, antisymmetric tensor and graviton of type $\si$.}
    \label{damour}
  \end{center}
\end{figure}

In \cite{Buonanno:1998kx} Buonanno and Damour also found a non-renormalization, 
but via a different cancellation. The authors of \cite{Buonanno:1998kx} analyzed a 
physical quantity which is described by a diagram of the 
type depicted in fig.~\ref{damour}, where it is understood that each of the 
fields interacting with the string can propagate in the internal line.
We now take a closer look at the different contributions to this process.
Letting a $\si_{ij}$ propagate in the wavy line of fig.~\ref{damour} yields a vanishing 
result given that the amplitude for such a process has the following behavior 
\be 
\label{null}
fig.~\ref{damour}_{\si_{ij}}\propto 
T^{ij}\rnode{g1}\sigma_{ij}\rnode{g2}\sigma_{kl}T^{kl}\propto\mu^2P_{11;11}=
0\,,
\ncbar[nodesep=2pt,angle=-90,armA=4pt,armB=4pt]{g1}{g2}
\ee
as it can be explicitly checked from eq.~(\ref{prop_ten}). This diagram 
vanishes for the same reason why two straight, static,
parallel strings do not exert a force on each other: the amplitude for one graviton 
exchange between two such strings is proportional to the same vanishing quantity 
$P_{11;11}$. 
The dilaton contribution to the amplitude of fig.~\ref{damour} is
\be
\label{ren}
fig.~\ref{damour}_\Phi=\frac{\alpha^2\mu^2}{2M_{Pl}^2}\int \frac {d^2k}{k^2}\,,
\ee
whereas to find the effect of the antisymmetric tensor it is enough to replace 
$\alpha^2$ with $-\beta^2$ in eq.(\ref{ren}), as can be checked using (\ref{prop_b}) and 
(\ref{gammazero}).
These three amplitudes, condensed in the representation of fig.~\ref{damour}, 
have a close correspondence with what is found in \cite{Buonanno:1998kx} and 
show that the contributions to the superstring renormalization are different when 
calculated by looking at the self-energy as in \cite{Buonanno:1998kx} 
other than through the (effective) EMT as in \cite{Dabholkar:1989jt} and in the present work; 
nonetheless, the non-renormalization property of superstrings is preserved in both approaches.

The source of the discrepancy is explained in \cite{Buonanno:1998kx}
where it is observed that the difference in the two ways of computing 
the renormalization of the string tension amounts to a (divergent) 
source-localized term, "as the interaction-energy cannot be unambiguously 
localized only in the field, there are also interaction-energy contributions 
which are localized in the sources" which are missed in one approach 
but accounted for in the other. 

Moreover, the contribution of the antisymmetric tensor to the string tension 
renormalization turns out to be the same with the two methods because this 
coupling to the string is metric-independent, so it does not contribute to the total 
EMT given by $T^{\mu\nu}\equiv 2 g^{-1/2} \delta S / \delta g_{\mu\nu}$.
Of course the physical result cannot depend on the details of the calculation method: 
indeed the source-localized contribution just renormalizes the \emph{bare} tension of 
the string and does not give physical effects.
As observed in \cite{Buonanno:1998kx}, this constrasts Dirac's argument 
\cite{Dirac} about the connection bewteen the renormalization of a point 
charge and its divergent field self-energy.


Therefore, we support the explanation of the discrepancy given by Buonanno 
and Damour \cite{Buonanno:1998kx} and provide a computation of the 
renormalization of the EMT with a completely different technique than in Dabholkar and 
Harvey \cite{Dabholkar:1989jt}, confirming their result.

Following the track of the EFT methodes we employed, one could also compute the 
renormalization of the couplings of $\Phi$ and 
$B_{\mu\nu}$. 
For the dilaton coupling the relevant diagrams are two, both of the type 
fig.~\ref{str_1PN}, with a $\Phi$ as outer wavy line and either two $B_{\mu\nu}$'s
or a $\Phi$ and a $\sigma_{ij}$ as dashed inner lines.
For the antisymmetric tensor case, the external $B_{\mu\nu}$ can be attached to 
either a $\si_{ij}$ and a $B_{\mu\nu}$ or to a $\Phi$ and a $B_{\mu\nu}$. 
All the above mentioned trilinear vertices have the same dependence on external 
momentum as the gravity case. 

One final remark is needed about result (\ref{eemt_s}). A tensor $T_{ab}(x)$ 
is conserved if $T_{ab}^{\phantom{ab},b}(x)=0$ which, in Fourier space, 
translates naively to
\be
\label{naive_conserv}
\dpa^aT_{ab}(q)\stackrel ?=-i q^a T_{ab}(q)\qquad {\rm NO!}
\ee
Clearly, with an EMT of the form (\ref{eemt_s}), for $\alpha^2\neq\beta^2$ 
the right hand side of eq.~(\ref{naive_conserv}) does not vanish. This happens 
because $T_{ab}(q)$ is not square integrable, thus it is not ensured that the 
derivative operation and the Fourier transform commute with each other, 
and indeed they do not in this case, see Appendix for details. 


\section{Conclusions}
\label{conclusion}

We have studied point-like and one-dimensional-extended sources in the context 
of scalar-tensor gravity and we have computed the effects of fields self-interactions
to the renorma\-lization of the effective energy-momentum tensor.

The calculations have been performed within the framework provided by the 
effective field theory methods applied to gravity \cite{Goldberger:2004jt,Goldberger:2007hy}, exploiting the powerful tool of a systematic expansion in terms of Feynman diagrams.

The classical ``dressing'' of the sources by long range interactions has the effect of 
smearing the source, consistently with coordinate covariance, and implies energy-momentum tensor conservation.
We obtained perturbative solutions valid to first post-Newtonian order or, in 
the case of extra scalar fields, up to the trilinear interaction terms.

In the case of a string source we reviewed the renormalization of both its 
effective energy-momentum tensor and its tension, which has
been subject of investigation with apparently conflicting results in the past 
\cite{Dabholkar:1989jt,Buonanno:1998kx}.
We exposed the fully satisfactory explanation of the discre\-pancy given by 
Buonanno and Damour \cite{Buonanno:1998kx} and confirmed that the 
renormalization of the energy-momentum tensor and the renormalization of the 
string tension differ by source-localized contributions.

\section*{Acknowledgements}

This work is supported by the Fonds National Suisse. The work of R.~S. is 
supported by the Boninchi foundation.
The authors wish to thank M.~Maggiore for discussions and support, 
W.~Goldberger for his prompt and helpful correspondence and R.~Porto for 
introducing them to Feyn Calc.
R.~S. wishes to thank A. Nicolis for useful discussions and the Aspen Center 
for Physics for organizing a very stimulating workshop on gravitational wave 
astronomy.

\appendix
\section{}

To second order the metric (\ref{met_nr}) can be rewritten as
\be
g_{\mu\nu}=\pa{
\ba{cc}
-1-2\varphi -2\varphi^2& a_j+2\varphi a_j \\
a_i+2\varphi a_i & \quad\delta_{ij}-2\varphi(\delta_{ij}+\varsigma_{ij})
+2\varphi^2\delta_{ij}+\varsigma_{ij}-a_ia_j\\
\ea
}\,,
\ee
where $\gamma_{ij}\equiv \delta_{ij}+\varsigma_{ij}$ (exact).
It is also useful to have the form of the inverse metric  
\be
g^{\mu\nu}=\pa{
\ba{cc}
-e^{-2\varphi}\pa{1-e^{4\varphi}\gamma^{ij}a_ia_j} & e^{2\varphi}a^j\\
e^{2\varphi}a^i & e^{2\phi}\gamma^{ij}\\
\ea
}\,.
\ee
To second order one has
\be
g^{\mu\nu}\simeq\pa{
\ba{cc}
-1+2\varphi-2\varphi^2+\delta_{ij}a_ia_j & a_j+2\varphi a_j-\varsigma_{jk}a_k\\
a_i+2\varphi a_i -\varsigma_{ik}a_k & 
\delta_{ij}+2\varphi\delta_{ij}-\varsigma_{ij}+
2\varphi\pa{\varphi\delta_{ij}-\varsigma_{ij}}\\
\ea
}\,.
\ee

The relevant integral for computing Feynman diagrams like the one represented 
in fig.~\ref{pp_1PN} (see for instance \cite{itz} and 
\cite{BjerrumBohr:2002ks}) is
\renewcommand{\arraystretch}{1.4}
\be
\label{kikj}
\ba{rcl}
\ds\int \frac{d^3\K}{\pa{2\pi}^3}\frac{k^ik^j}{k^2(\K+\Q)^2} &=&\ds
\frac 1{64}\pa{-\delta^{ij}q+3\frac{q^iq^j}q}\,,\\
\ds\int \frac{d^3\K}{\pa{2\pi}^3}\frac{k^i}{k^2(\K+\Q)^2}&=&
\ds\frac{q^i}{16q}\,.
\ea
\ee
The integral relevant for fig.~\ref{3psi} is
\be
\int \frac{d^3\K}{\pa{2\pi}^3}\frac 1{(k^2+M^2)\paq{(\K+\Q)^2+M^2}}=
\frac {1}{4\pi q}\arctan\pa{\frac q{2M}}\,,
\ee
and to reconstruct the metric out of the effective EMT we used
\be
\label{int_k}
\ba{lcl}
\ds\int \frac{d^3\Q}{\pa{2\pi}^3}\,e^{i\Q\cdot{\bf x}}\,\frac 1{q^2}&=&
\ds\frac 1{4\pi |{\bf x}|}\,,\\
\ds\int \frac{d^3\Q}{\pa{2\pi}^3}\,e^{i\Q\cdot{\bf x}}\,\frac 1{q}&=&\ds
\frac 1{2\pi^2 |{\bf x}|^2}\,,\\
\ds\int \frac{d^3\Q}{\pa{2\pi}^3}\,e^{i\Q\cdot{\bf x}}\,\frac {q^iq^j}{q^3}&=&
\ds\frac 1{2\pi^2 |{\bf x}|^2}\pa{\delta_{ij}-2\frac{x^ix^j}{|{\bf x}|^2}}\,.
\ea
\ee

The relevant integral for computing Feynman diagrams like the one represented 
in fig.~\ref{str_1PN} are (see again \cite{itz})
\be
\label{kikj_2}
\ba{rcl}
\ds\int \frac{d^{2+\eps}k}{\pa{2\pi}^{2+\eps}}\frac{k^ik^j}{k^2(k+q)^2}&=&
\ds\frac {\pa{q^2}^{\eps/2}}{\pa{4\pi}^{1+\frac \eps 2}}\left[
\frac 12\delta^{ij}\Gamma\pa{-\frac \eps 2}\int_0^1 
\paq{x(1-x)}^{\frac \eps 2}dx+\right.\\
&&\ds\qquad\qquad\left.\,\frac{q^iq^j}{q^2}\Gamma\pa{1-\frac\eps 2}
\int_0^1x^2\paq{x(1-x)}^{\frac\eps 2-1}dx\right]\,,\\
\ds\int \frac{d^{2+\eps}k}{\pa{2\pi}^{2+\eps}}\frac{k^i}{k^2(k+q)^2} &=&
\ds\frac{q^i}{\pa{q^2}^{1-\eps/2}\pa{4\pi}^{1+\frac\eps 2}}
\Gamma\pa{1-\frac\eps 2}\int_0^1 x\paq{x\pa{1-x}}^{\frac \eps 2-1}dx\,.
\ea
\ee

Other useful formulas to anti-Fourier transform the string effective EMT at 
next-to-leading order, are
\be
\int \frac{d^2q}{2\pi^2} \log(q) e^{iqx} &=&-\frac 1{x^2}\\
\int_{q_\eps} \frac{d^2q}{\pa{2\pi}^2}\frac 1{q^2}e^{iqx}&=&
-\frac 1{2\pi}\log(xq_\eps)+\frac{\ln 2-\gamma}{2\pi}+\frac{r^2q_\eps^2}{16\pi}
+o\paq{\pa{xq_\eps}^3}\,,
\ee
where a disk of radius $q_\eps$ around the origin has been cut out of the 
integral. Moreover  
\be
\int_0^{2\pi}e^{ix\cos\theta}d\theta=2\pi J_0(x)\,,
\ee
where $J_0$ is the Bessel function of zero-th order.
To derive the metric out of the string effective EMT the following integral
\be
\int_{q_\eps} \frac{d^2q}{\pa{2\pi}^2}\frac 1{q^4}e^{iqx}=
\frac{x^2}{2\pi}\paq{\frac 1{2q_\eps^2r^2}+\frac 18\log\pa{q_\eps^2r^2}+
\frac{\gamma-\ln 2-1}4+o(xq_\eps)}
\ee
is helpful.

The effective EMT (\ref{eemt_s}) in coordinate space is
\be
\label{str_eemtI}
T_{ij}^{(I)}(x)=-\frac 4\pi G_N\mu^2\pa{\ba{cc}
\pa{\alpha^2-\beta^2}\pa{C'\delta^{(2)}(x^a)+1/r^2}& \\
&\ds\frac{\alpha^2+\beta^2}{r^2}\pa{-\frac 12\delta_{ab}+\frac{x_ax_b}{r^2}}
\ea}\,,
\ee
where $r$ denotes the distance to the string in the transverse two-dimensional
space. Here $C'$ denotes the $q$-independent part of the quantity defined in 
text in (\ref{div_const}).

To explicitly check conservation in the Fourier space of the string effective
EMT (\ref{eemt_s}), let us write down the conservation of the EMT in $q$-space,
keeping only the components transverse to the string world-sheet:
\be
\label{c_emt}
\dpa^aT_{ab}(q)=\int d^2x\paq{\dpa^a T_{ab}(x)}e^{iqx}=
\int d^2x\paq{\dpa^a\pa{T_{ab}e^{iqx}}-iq^aT_{ab}(q)
e^{iqx}}\,,
\ee
which has an extra piece with respect to (\ref{naive_conserv}).
Let us restrict for simplicity to the total derivative term and let us fix 
the index $b=2$.
To make sense of the integral we have to integrate over a region $\Omega$ 
obtained by cutting out of the plane the the two regions $r<r_\eps$ and $r>R$, 
and we will finally (but after taking the other limits first) let 
$r_\eps\to 0$ and $R\to\infty$.\\
By changing coordinates from $y,z$ to $\rho,\theta$ according to 
$y=r\cos\theta$, $z=r\sin\theta$ and using the Green-Gauss theorem
one obtains
\be
\ba{rl}
\ds -\frac{\pi}{4G_N\mu^2\pa{\alpha^2+\beta^2}}
\int_\Omega d^2x &\dpa^a\paq{T_{a2}(x)e^{iq_ax^a}}=
\ds \int_{\dpa\Omega}\frac 1{2r}\cos\theta e^{iqr\cos\theta} d\theta\\
=&\ds\frac 1{2R}\int_0^{2\pi}\cos\theta e^{iqR\cos\theta}d\theta -
\frac 1{2r_\eps}\int_0^{2\pi} \cos\theta e^{iqr_\eps\cos\theta}d\theta\,.
\ea
\ee
The first integral is clearly vanishing in the limit $R\to\infty$. Expanding 
the exponential in the second integral, taking the limit $r_\eps\to 0$ and
finally plugging this result into (\ref{c_emt}), one has
\bdm
\dpa^aT_{ab}(q)\propto \frac{iq^a}2-
iq^a\pa{-\frac{\delta_{ab}}2+\frac{q_aq_b}{q^2}}=0\,,
\edm 
qed.

\end{document}